\documentclass{article}
\usepackage{graphicx} 
\usepackage{booktabs}
\usepackage{siunitx}
\usepackage{authblk}
\title{RoMedFormer: A Rotary-Embedding Transformer Foundation Model for 3D Genito-Pelvic Structure Segmentation in MRI and CT}
\author{
    Yuheng Li\textsuperscript{1,2}\thanks{Co-first authors}, 
    Mingzhe Hu\textsuperscript{1,3}\thanks{Co-first authors}, 
    Richard L.J. Qiu\textsuperscript{1}, 
    Maria Thor\textsuperscript{4}, 
    Andre Williams\textsuperscript{5}, 
    Deborah Marshall\textsuperscript{5}, 
    Xiaofeng Yang\textsuperscript{1,2,3}\thanks{Corresponding author: \texttt{xiaofeng.yang@emory.edu}}  
}

\date{}

\affil{\small \textsuperscript{1}Department of Radiation Oncology and Winship Cancer Institute, Emory University, Atlanta, GA 30308}
\affil{\small \textsuperscript{2}Department of Biomedical Engineering, Emory University, Atlanta, GA 30308}
\affil{\small \textsuperscript{3}Department of Computer Science and Informatics, Emory University, Atlanta, GA 30322}
\affil{\small \textsuperscript{4}Department of Medical Physics, Memorial Sloan Kettering Cancer Center, New York, NY 10065}
\affil{\small \textsuperscript{5}Department of Radiation Oncology, Icahn School of Medicine at Mount Sinai, New York, NY 10029}

\begin{document}

\maketitle
\vspace{-1cm}
\begin{abstract}
Deep learning-based segmentation of genito-pelvic structures in MRI and CT is crucial for applications such as radiation therapy, surgical planning, and disease diagnosis. However, existing segmentation models often struggle with generalizability across imaging modalities, and anatomical variations. In this work, we propose \textbf{RoMedFormer}, a rotary-embedding transformer-based foundation model designed for 3D female genito-pelvic structure segmentation in both MRI and CT. RoMedFormer leverages self-supervised learning and rotary positional embeddings to enhance spatial feature representation and capture long-range dependencies in 3D medical data. We pre-train our model using a diverse dataset of 3D MRI and CT scans and fine-tune it for downstream segmentation tasks. Experimental results demonstrate that RoMedFormer achieves superior performance segmenting genito-pelvic organs. Our findings highlight the potential of transformer-based architectures in medical image segmentation and pave the way for more transferable segmentation frameworks.
\end{abstract}

\section{Introduction}
Gynecologic and other pelvic cancers affect millions of women worldwide each year,\cite{bray2024global} with radiation therapy (RT) remaining a central treatment modality. While RT has significantly improved survival rates, it also poses long-term risks to critical genito-pelvic structures, including the vagina, bulboclitoris, labia, and associated neurovascular networks. Radiation-induced damage to these structures can result in severe complications such as vaginal stenosis, dyspareunia, orgasm dysfunction, loss of sexual sensation \cite{marshall2022female}, and overall pelvic dysfunction, which greatly impact post-treatment quality of life \cite{basch2023patient, lindau2015manifesto}. Despite these risks, current RT planning protocols have historically focused on protecting male sexual and reproductive organs—such as the prostate and penile structures, while female-specific anatomy has been largely overlooked \cite{unger2022sex, peters2023clinical}. This disparity stems from long-standing biases in medical imaging datasets and clinical segmentation practices, which have led to a critical gap in the accurate delineation and preservation of female genito-pelvic structures during RT planning.

One major challenge in addressing this issue is the lack of robust segmentation methods tailored to the female genito-pelvic anatomy. Manual segmentation, the current gold standard, is not only highly labor-intensive and time-consuming, but also prone to inter- and intra-observer variability, especially when delineating small, complex, and interconnected anatomical structures \cite{jannini2014beyond} such as the bulboclitoris, greater vestibular and lesser vestibular (paraurethral) glands, and vaginal wall muscular and fibrovascular components. These structures exhibit subtle contrast variations in CT and MRI scans, making them particularly challenging to delineate consistently, even for experienced clinicians. Furthermore, the scarcity of annotated medical imaging datasets exacerbates this problem, as female-specific genito-pelvic structures are often missing or underrepresented in widely used segmentation datasets. The combination of anatomical complexity, dataset imbalance, and clinical bias creates a major barrier to optimizing RT planning for female patients \cite{di2014anatomic, peters2023clinical, doo2025sex}. To bridge this gap, there is an urgent need for advanced computational methods that can improve segmentation accuracy, consistency, and efficiency in the delineation of genito-pelvic structures.

The field of medical image segmentation has witnessed rapid advancements with the emergence of deep learning, evolving from convolutional neural network (CNN)-based approaches, such as U-Net \cite{ronneberger2015u} and nnU-Net \cite{isensee2021nnu}, to transformer-based architectures, including Swin-UNETR \cite{hatamizadeh2021swin} and TransUNet \cite{chen2021transunet}. These state-of-the-art (SOTA) models have demonstrated superior performance in segmenting complex anatomical structures by leveraging global contextual information through self-attention mechanisms. Despite these advancements, segmentation models remain disproportionately focused on well-documented anatomical structures, while genito-pelvic organs remain notably underrepresented in contemporary AI-driven segmentation research. This disparity can be attributed to bias in dataset curation and the scarcity of annotated genito-pelvic medical imaging data, both of which hinder the development of robust AI models for female pelvic anatomy. Notably, over 10\% of FDA-approved AI-based segmentation tools are designed for prostate cancer, yet none explicitly address female pelvic structures, reflecting a broader imbalance in medical AI task/tool development. Moreover, commercial AI-based segmentation software exhibits similar biases, with existing solutions predominantly tailored to the male pelvis while lacking dedicated models for female genito-pelvic structures. The absence of such models in both regulatory-approved and commercial AI offerings further limits the availability of automated tools that can support accurate delineation of female pelvic organs. More equitable and generalizable AI-driven segmentation models that can address the limitations of existing approaches are urgently needed for female patients.

Large foundation models (LFMs) have transformed medical image segmentation by enabling scalable and generalizable learning across diverse imaging tasks \cite{wang2025triad, ma2024segment, wang2022medclip, li2025automatic}. However, existing models primarily focus on well-represented anatomical structures, with genito-pelvic organs remaining critically underrepresented. We introduce RoMedFormer, the first foundation model adapted for genito-pelvic structure segmentation. 1. Unlike existing LFMs that emphasize common anatomical regions, RoMedFormer is tailored to delineate key female pelvic structures. 2. We employ a multi-stage training pipeline, beginning with self-supervised pretraining on large-scale unlabeled datasets, followed by supervised fine-tuning on TotalSegmentator \cite{wasserthal2023totalsegmentator} and AMOS22 \cite{ji2022amos} datasets and task-specific adaptation to genito-pelvic segmentation, ensuring improved generalization despite data scarcity. 3. RoMedFormer is a multimodality model, supporting segmentation across both MRI and CT scans, enhancing its adaptability to different clinical imaging modalities. 4. We optimize the transformer architecture by Rotary Positional Embeddings \cite{heo2024rotary} for improved spatial representation, and Swish-Activation Gated Linear Units \cite{grattafiori2024llama} for enhanced feature learning, while employing a lightweight convolutional decoder to reduce computational overhead. These advancements establish RoMedFormer as a robust and efficient foundation model for genito-pelvic structures segmentation radiation therapy planning.

\section{Method}
\subsection{Data acquisition}
Our genito-pelvic dataset is derived from the Sexual Toxicity After RT (STAR) Study, which enrolls female patients receiving pelvic RT for non-metastatic anal, vaginal, vulvar, rectal, cervical, and uterine cancers. 

Imaging data consists of scans from multiple modalities, including CT and MRI, collected for target and organ delineations, treatment planning and anatomical assessments. CT scans were acquired during CT-based RT simulation. These CT scans have a spatial resolution of 1.6 × 1.6 × 2 mm. MRI scans were obtained following trial imainging protocol using T2-weighted sequences with a resolution of 0.5 × 0.5 × 2 mm, ensuring high soft-tissue contrast for the delineation of genito-pelvic structures. Our dataset includes a total of 30 imaging studies, comprising 10 RT planning CT scans and 20 T2-weighted MRI scans.

For segmentation, all images were manually contoured and approved by at least two expert radiation oncologists, serving as the ground truth contours for model training and validation. Data anonymization was performed using the MIM Anonymizer, and contours were verified through a standardized workflow incorporating rigid and deformable image registration. 

\subsection{Multi-Stage Learning Pipeline}
To effectively segment genito-pelvic structures while addressing data scarcity and anatomical complexity, we propose a multi-stage learning strategy that progressively refines the model from general anatomical understanding to highly specialized segmentation. Our approach (shown as Figure 1) consists of self-supervised pretraining, supervised fine-tuning, and task-specific fine-tuning, ensuring both broad generalization and precise adaptation to genito-pelvic anatomy.

\begin{figure}[htbp] 
    \centering
    \includegraphics[width=\textwidth]{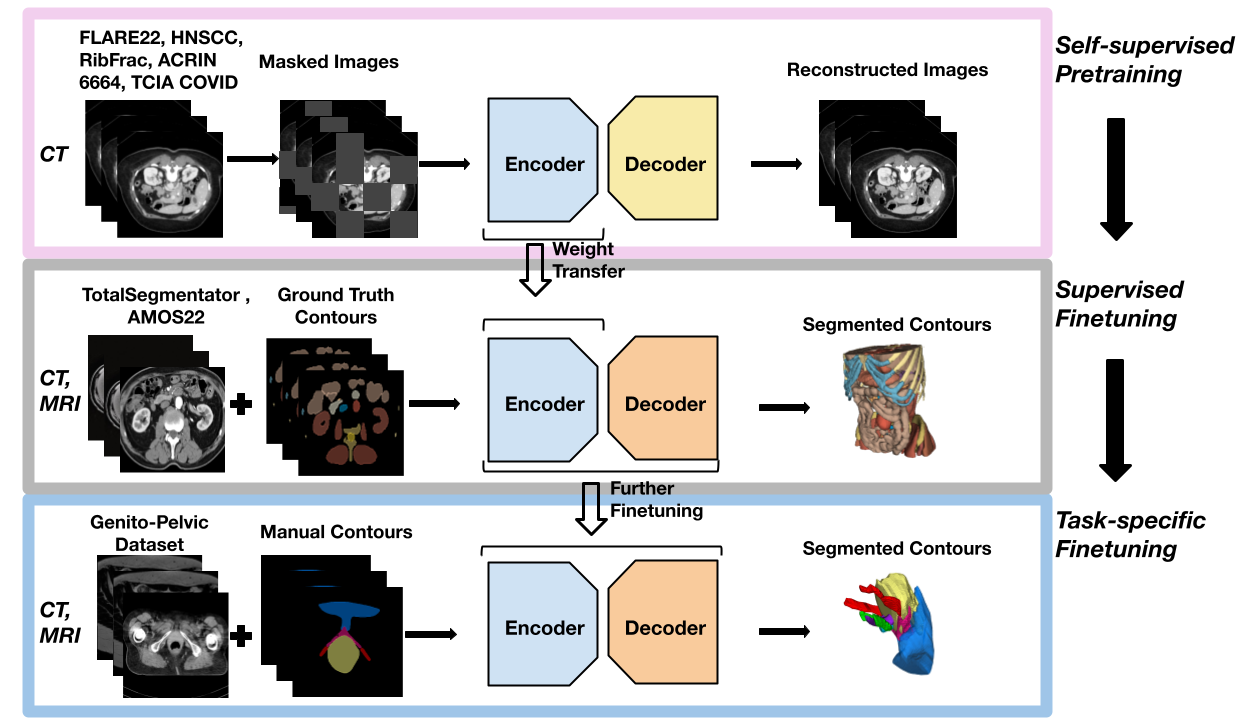} 
    \caption{Overview of the Multi-Stage Learning Strategy for Genito-Pelvic Structure Segmentation. The model undergoes three progressive training stages: (1) Self-supervised pretraining, where it learns general anatomical representations from large-scale CT datasets using masked image modeling; (2) Supervised fine-tuning, leveraging annotated CT dataset such as TotalSegmentator and cross-modality CT-MRI dataset AMOS22 to refine segmentation accuracy across multiple organ systems; and (3) Task-specific fine-tuning, where the model is further adapted to genito-pelvic segmentation using a dedicated dataset. Weight transfer between stages enables efficient feature learning and adaptation.}
    \label{fig:pipeline}
\end{figure}

In the first stage, self-supervised pretraining is performed on a large collection of CT datasets, including FLARE22 \cite{ma2022fast}, HNSCC \cite{grossberg2017data}, RibFrac \cite{jin2020deep}, ACRIN 6664\cite{smith2015data}, and TCIA COVID \cite{an2020ct}, allowing the model to learn foundational anatomical representations without requiring manual annotations. This stage utilizes masked image modeling (MIM), where parts of the image are intentionally hidden, and the model is trained to reconstruct them, forcing it to develop a deep contextual understanding of anatomical structures. This approach helps extract meaningful feature representations that generalize across different imaging domains.

The second stage, supervised fine-tuning, builds upon the learned representations by training on TotalSegmentator \cite{wasserthal2023totalsegmentator}, a comprehensive multi-organ segmentation dataset with annotations for 104 anatomical structures. The transformer is further trained on multimodality AMOS22 \cite{ji2022amos} which contains both CT and MRI data by transferring weights from TotalSegmentator. The Transformer encoder is initialized from the pretraining phase to retain global anatomical features, while the decoder is trained to specialize in multi-organ segmentation tasks, enhancing spatial precision and region-specific accuracy.

The final stage, task-specific fine-tuning, is dedicated to genito-pelvic structure segmentation, using an institutional dataset, consisting of MRI and CT scans of patients who opted for receiving pelvic RT. This stage ensures that the model adapts to the unique characteristics of female genito-pelvic anatomy, including soft tissue structures with low contrast, small neurovascular features, and high inter-patient variability. Fine-tuning at this stage allows the model to achieve highly specialized segmentation performance tailored to clinical needs.

\subsection{Model architecture}
\textbf{Motivation.} Our motivation stems from the limitations of previous segmentation methods, which predominantly rely on large or highly specialized convolutional neural networks \cite{huang2023stu}. We posit that Transformer architectures offer significant advantages for medical image segmentation, primarily due to their inherent sequence modeling capabilities. Firstly, healthcare is inherently multimodal, making the seamless integration of multiple data modalities essential. Representing high-dimensional 3D medical images as token sequences simplifies incorporating visual information alongside other data domains, facilitating tasks such as generating precise clinical reports from CT or X-Ray scans \cite{hamamci2024developing, stock2024generalist}. Secondly, transformer-based tokenization enhances computational efficiency, particularly when employed within masked image modeling frameworks for self-supervised learning (SSL). Consequently, transformers have become the preferred architecture in SSL paradigms \cite{darcet2025cluster, oquab2023dinov2, peng2022beit}. 

\begin{figure}[htbp] 
    \centering
    \includegraphics[width=0.7\textwidth]{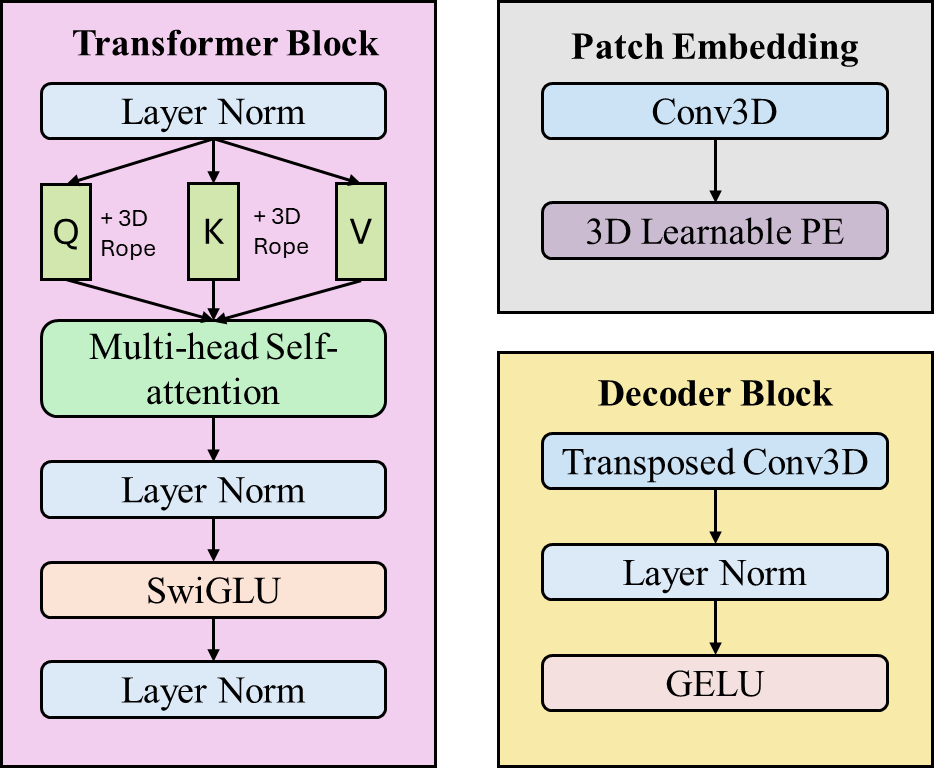} 
    \caption{Overview of our segmentation model design.}
    \label{fig:model}
\end{figure}

\textbf{3D Patch Embedding and Rotatory Embeddings.} To effectively leverage the transformer paradigm for novel structure segmentation in 3D medical images, we propose converting volumetric data into a 1D sequence of visual tokens. Traditional methods utilize tokenization with large patch sizes, typically 16 × 16 × 16, which significantly compresses the image data, potentially losing essential spatial information. Inspired by previous studies \cite{hatamizadeh2022unetr, wald2025primus}, we propose using a smaller patch size of 8 × 8 × 8. This reduced patch size allows us to preserve more fine-grained anatomical details critical for accurate segmentation. Another significant challenge in adapting transformers to medical imaging is handling spatial relationships, which are crucial in accurately segmenting anatomical structures. Standard transformers rely on positional embeddings to encode spatial information; however, absolute positional embeddings (APEs) may not effectively capture spatial relationships in medical images processed through sliding-window methods, which can introduce positional ambiguity due to patch-level shifts. To overcome this limitation, we adopt Rotary Positional Embeddings (RoPE) \cite{heo2024rotary}. RoPE effectively encodes relative spatial distances and orientations directly within the self-attention mechanism, making it more robust to capture the spatial correlations inherent in anatomical structures. This positional encoding strategy considerably enhances positional awareness, particularly beneficial when using patch-based processing for large 3D volumes.

\textbf{Block Design.} Motivated by advancements in natural language processing, we integrate Swish-Activation Gated Linear Units (SwiGLUs) into our Transformer architecture \cite{grattafiori2024llama}. SwiGLUs have demonstrated improved representational power compared to standard MLP blocks used in Vision Transformers (ViT) \cite{oquab2023dinov2}. To further reduce computational complexity, we employ a lightweight convolutional decoder consisting of sequential transposed convolution, normalization, and activation blocks. This minimal yet effective decoder architecture reconstructs the full-resolution image from the token sequence efficiently, ensuring lower computational overhead compared to deeper convolutional alternatives such as U-Net. A detailed overview of our block designs is shown in Figure \ref{fig:model}. 

\subsection{Experiments and implementation details}
We preprocessed the dataset using Residual Encoder L U-Net (ResEnc-L) plan. The patch size is 48 $\times$ 320 $\times$ 320, with spacing of 1.9 mm $\times$ 0.5 mm $\times$ 0.5 mm. A five-fold cross-validation was performed on our genito-pelvic dataset. We used a training/testing ratio of 0.8/0.2, with an average fold consisting of 14 volumes for training and 3 volumes for testing. We use the nnU-Net framework with each fold being trained for 1000 epochs with 250 steps per epoch. The hyperparameters are as follows: batch size 2, learning rate 3e-4, weight decay 5e-2, AdamW optimizer and gradient clipping of 1. The drop ratio of DropPath was set to 0.2 and Layer Scale was initialized with 0.1. The loss is DiceCE. 

\subsection{Evaluation metrics}
We employed four metrics to quantitatively assess the performance of RoMedFormer:

\begin{itemize}
    \item \textbf{Dice Coefficient}: Measures the overlap between our segmentations and ground truth segmentations.
    \item \textbf{Volume Difference (VolDiff)}: Evaluates the volumetric accuracy between our segmentations and references in cubic centimeters (cc).
    \item \textbf{Average Surface Distance (SurfDist)}: Captures the average surface discrepancy between our segmentations and ground truth surfaces.
    \item \textbf{Centroid Distance (CentroidDist)}: Measures the Euclidean distance between the centroids of our segmentations and reference structures.
\end{itemize}

\section{Result}
\subsection{Qualitative results}
\begin{figure}[htbp] 
    \centering
    \includegraphics[width=\textwidth]{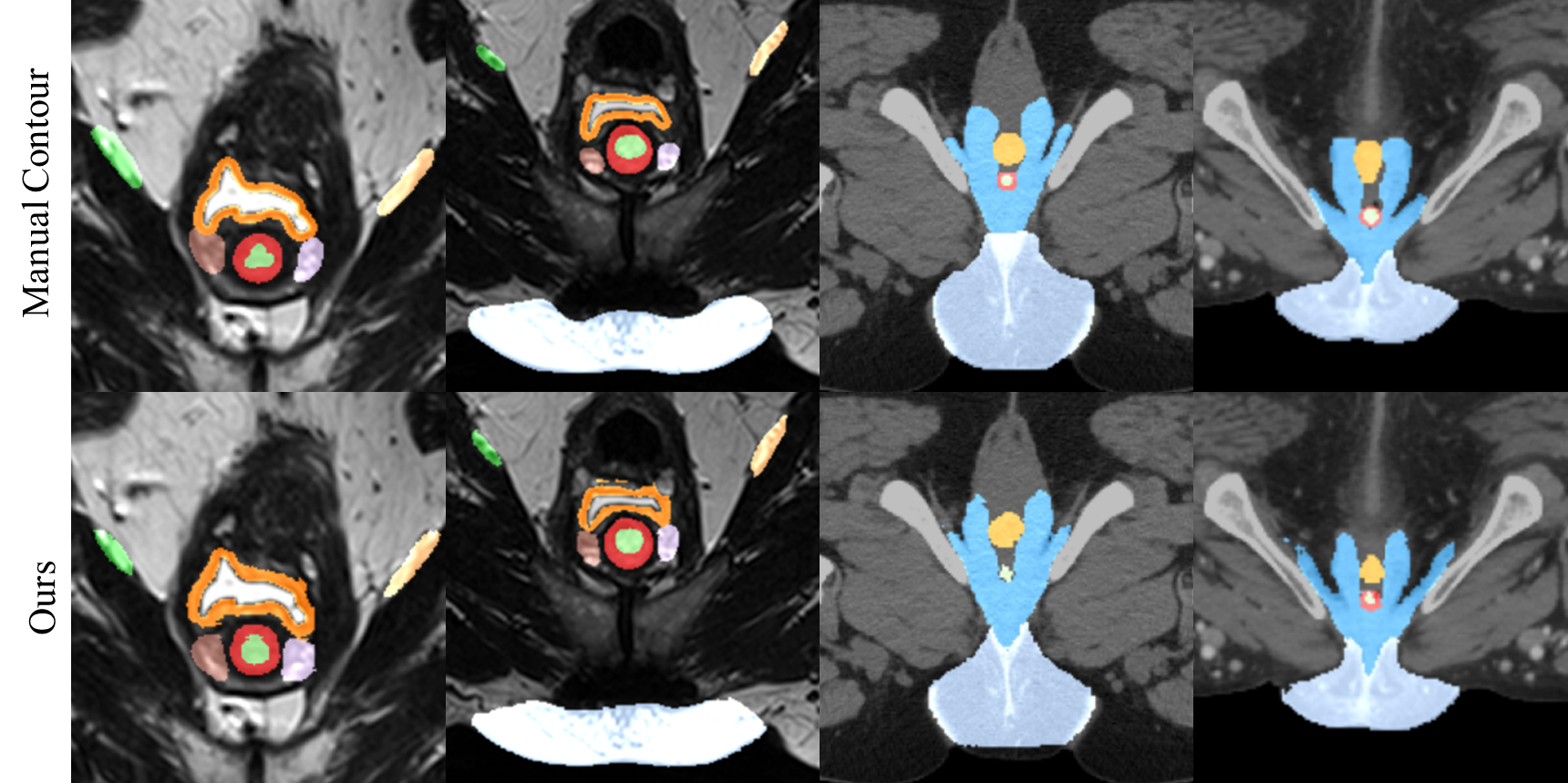} 
    \caption{Visualizations of segmentation results on GYN. Column 1 and 2 shows model segmentations on MRI images and column 3 and 4 shows segmentations on CT. Dark Blue: Bulboclitoris. Light Blue: Genitals. Orange: Left Neurovascular Bundle (Internal Pudendal). Green: Right Neurovascular Bundle (Internal Pudendal). Light Green: Urethra. Red: Paraurethral Gland. Pink: Left Ovary. Light Pink: Right Ovary. Purple: Left Neurovascular Bundle (Inferior Hypogastric). Brown: Right Neurovascular Bundle (Inferior Hypogastric).}
    \label{fig:qualitative}
\end{figure}
We demonstrate RoMedFormer's segmentation capabilities in Figure \ref{fig:qualitative}. We found that RoMedFormer's segmentations closely match the ground truth across MRI and CT modalities, especially for the genitals and neurovascular bundles, with consistent shape and boundary correspondence to the reference annotations. However, we still observe minor segmentation artifacts due to the small training sample size. We believe collecting a large-scale female genito-pelvic dataset can further improve the model's ability to segment anatomically challenging structures.

\subsection{Quantitative results}
  Quantitative results for RoMedFormer's segmentation performance across different anatomical structures are summarized in Table \ref{tab:metrics}, which includes data from 6 CT scans and 11 MRI scans. RoMedFormer achieved an overall Dice coefficient of 0.7060 ± 0.0762 across all classes, indicating strong segmentation accuracy. Among individual structures, the best performance was observed in segmenting the genitals (Dice: 0.8677 ± 0.0219), followed by the bulboclitoris (Dice: 0.8232 ± 0.0451). Segmentation of the vagina was more challenging, yielding a lower Dice coefficient of 0.5668 ± 0.1643. In terms of volumetric accuracy, VolDiff across all structures was -0.4301 ± 2.2531 cc, reflecting minimal bias in our segmentations. Notably, the right ovary exhibited the highest negative deviation (-2.1723 ± 5.2636 cc), whereas the left ovary had the highest positive deviation (2.7775 ± 2.4767 cc), indicating variability in accurately predicting ovarian volumes. Surface distance analysis revealed an average distance of 0.5743 ± 0.2841 mm across all structures. The smallest average surface distances were observed in the urethra (0.2024 ± 0.0295 mm) and the bulboclitoris (0.2813 ± 0.0833 mm), demonstrating precise boundary delineation. In contrast, the left ovary showed the greatest average surface distance (1.1711 ± 0.6128 mm), reflecting segmentation boundary uncertainty in this region. Lastly, RoMedFormer achieved an average centroid distance (CentroidDist) of 3.6893 ± 1.8465 mm across all structures. The paraurethral gland and urethra showed the smallest centroid deviations (1.6335 ± 0.0697 mm and 1.6196 ± 0.9891 mm, respectively), signifying accurate localization. Conversely, the left ovary demonstrated the highest centroid deviation (6.4334 ± 2.2691 mm), highlighting localization challenges for this structure. These quantitative metrics illustrate RoMedFormer's overall capability and areas of variability in segmentation accuracy, emphasizing both its strengths and the potential for further improvement in anatomically challenging structures.

\begin{table}[h]
    \centering
    \caption{Quantitative performance for each class, showing mean $\pm$ standard deviation.}
    \renewcommand{\arraystretch}{1.2}
    \resizebox{\textwidth}{!}{%
    \begin{tabular}{l c c c c}
        \hline
        \textbf{Organ class} & \textbf{Dice} & \textbf{VolDiff (cc)} & \textbf{SurfDist (mm)} & \textbf{CentroidDist (mm)} \\
        \hline
        Bulboclitoris           & $0.8232 \pm 0.0451$    & $1.3014 \pm 0.4866$      & $0.2813 \pm 0.0833$     & $2.4602 \pm 1.0084$ \\
        Genitals              & $0.8677 \pm 0.0219$    & $0.0376 \pm 6.3152$      & $0.2417 \pm 0.0822$     & $2.9187 \pm 2.2238$ \\
        Vagina                & $0.5668 \pm 0.1643$    & $-1.4817 \pm 1.5032$     & $0.6356 \pm 0.3726$     & $3.5234 \pm 1.7144$ \\
        Left Neurovascular Bundle (Internal Pudendal)  & $0.6868 \pm 0.0622$    & $-0.0798 \pm 1.5334$     & $0.4255 \pm 0.1452$     & $4.9280 \pm 4.3574$ \\
        Right Neurovascular Bundle (Internal Pudendal) & $0.6818 \pm 0.0365$    & $-0.6140 \pm 0.7388$     & $0.5292 \pm 0.0987$     & $2.3177 \pm 1.4990$ \\
        Urethra               & $0.7727 \pm 0.0548$    & $0.1084 \pm 0.1842$      & $0.2024 \pm 0.0295$     & $1.6196 \pm 0.9891$ \\
        Paraurethral Gland    & $0.7324 \pm 0.0110$    & $0.1012 \pm 0.5622$      & $0.2440 \pm 0.0343$     & $1.6335 \pm 0.0697$ \\
        Left Ovary            & $0.6957 \pm 0.0771$    & $2.7775 \pm 2.4767$      & $1.1711 \pm 0.6128$     & $6.4334 \pm 2.2691$ \\
        Right Ovary           & $0.6778 \pm 0.2021$    & $-2.1723 \pm 5.2636$     & $0.9070 \pm 0.8736$     & $4.4247 \pm 2.5339$ \\
        Left Neurovascular Bundle (Inferior Hypogastric)  & $0.6539 \pm 0.0680$    & $-2.4581 \pm 3.5373$     & $0.6945 \pm 0.2943$     & $4.5519 \pm 2.0924$ \\
        Right Neurovascular Bundle (Inferior Hypogastric) & $0.6070 \pm 0.0947$    & $-2.2517 \pm 2.1823$     & $0.9854 \pm 0.4991$     & $5.7715 \pm 1.5538$ \\
        \hline
        \textbf{Classwise Average} & $0.7060 \pm 0.0762$    & $-0.4301 \pm 2.2531$    & $0.5743 \pm 0.2841$     & $3.6893 \pm 1.8465$ \\
        \hline
    \end{tabular}
    }
    \label{tab:metrics}
\end{table}

\section{Discussion}
\textbf{Performance Analysis Across Genito-Pelvic Structures}
Our experimental results demonstrate that RoMedFormer effectively segments genito-pelvic structures across both MRI and CT modalities, highlighting its adaptability to multimodality imaging. However, segmentation performance varies depending on imaging contrast, organ size, and anatomical complexity. Larger, well-contrasted structures such as the genitals and clitoris achieved the highest Dice scores, suggesting that the model benefits from their distinct anatomical boundaries and consistent visibility across imaging modalities. Similarly, the urethra and paraurethral gland exhibited stable segmentation accuracy, likely due to their clearly defined anatomical boundaries and moderate contrast in both MRI and CT scans. In contrast, smaller and more complex structures, such as the inferior hypogastric neurovascular bundles and the vagina, demonstrated lower Dice scores and higher centroid distance errors, suggesting that these structures are more challenging to delineate due to their subtle contrast variations and intricate spatial relationships with surrounding tissues. Additionally, segmentation performance differences between MRI and CT scans indicate that modality-dependent contrast variations influence the model’s ability to distinguish certain structures, with MRI generally providing better soft-tissue contrast for smaller neurovascular and muscular structures. Despite these challenges, the classwise average Dice score of 0.7060 underscores the effectiveness of RoMedFormer in segmenting diverse genito-pelvic structures across imaging modalities.

\textbf{Training Strategies for Multimodality Learning}
Our study adopts a multimodality learning approach, where MRI and CT scans are combined during training and testing to enhance the generalizability of RoMedFormer across different imaging modalities. This strategy allows the model to leverage shared anatomical features across modalities, improving segmentation robustness despite inherent differences in image contrast and resolution. However, treating MRI and CT as interchangeable during training may limit the model’s ability to exploit modality-specific advantages, particularly given MRI’s superior soft-tissue contrast for neurovascular structures and CT’s clearer depiction of bony anatomy. A more advanced approach could involve modality-aware learning strategies, where the model explicitly learns to distinguish and integrate complementary information from both modalities.
One promising direction is cross-modality synthesis, where CT-to-MRI or MRI-to-CT translation using generative models could be employed to create synthetic paired datasets, allowing the model to learn richer representations by fusing modality-specific information. Another alternative is modality-specific feature fusion, where separate MRI- and CT-specific encoders extract independent feature embeddings that are later fused through cross-attention mechanisms or contrastive learning frameworks. Such an approach could help preserve modality-specific advantages while enhancing feature consistency across imaging domains. Future research could explore these modality-aware fusion techniques to further refine genito-pelvic structure segmentation, particularly for patients with access to both MRI and CT scans.

\textbf{Building a Benchmark Dataset for Genito-Pelvic Segmentation}
RoMedFormer is designed to achieve high segmentation performance even with limited training data, making it particularly suitable for anatomically complex structures where large annotated datasets are scarce. In this study, our dataset consisted of limited patients, and not all patients had both MRI and CT scans available, posing challenges for fully leveraging multimodality learning. While our current dataset provided valuable insights into model generalization across imaging modalities, a larger and more diverse dataset would enable further validation and refinement. We are actively working on expanding this dataset into a benchmark for genito-pelvic structure segmentation. This dataset will consist of additional MRI and CT scans from patients undergoing pelvic radiotherapy for non-metastatic anal, vaginal, vulvar, rectal, cervical, and uterine cancers, ensuring broader representation of anatomical variability. Imaging protocols will be standardized to include T2-weighted MRI, diffusion-weighted imaging, and dynamic contrast-enhanced MRI, along with high-resolution CT scans where available. Expert-validated anatomical contours will be incorporated to refine segmentation accuracy, and all data will be de-identified before processing.

\textbf{Integration of Clinical Text Information}
Incorporating clinical text information alongside imaging data can enhance genito-pelvic segmentation by providing anatomical and treatment-related context that is often challenging to infer from imaging alone. Radiology reports describe tumor presence, organ displacement, and imaging artifacts, offering insights into structures like the bulboclitoris, vaginal wall, and neurovascular bundles, especially in cases with poor soft-tissue contrast. Treatment planning records detail radiation dose distributions and organ-at-risk constraints, helping refine segmentation in radiosensitive regions. Surgical history and prior radiation exposure also play a critical role, as procedures like hysterectomy or pelvic exenteration alter anatomical landmarks. Additionally, structured text fields from electronic health records, including ECOG performance status and tumor staging, can guide model predictions by accounting for patient-specific variations. Future advancements could leverage vision-language models \cite{li2025towards, wang2022medclip, hu2024advancing} to align text-derived anatomical insights with imaging features, improving segmentation accuracy, particularly for complex or low-contrast structures.

\section{Conclusion}
In this work, we introduced RoMedFormer, the first large foundation model  for genito-pelvic structure segmentation in medical imaging. Unlike existing segmentation models that predominantly focus on well-represented anatomical structures, RoMedFormer is tailored to address the challenges associated with female genito-pelvic anatomy by incorporating domain-specific knowledge and multimodality adaptability. Through a multi-stage training pipeline, leveraging self-supervised pretraining, supervised fine-tuning, and task-specific adaptation, our model effectively overcomes data scarcity while enhancing generalization across diverse imaging datasets. Our results demonstrate the capability of RoMedFormer to segment across a wide range of genito-pelvic structures. RoMedFormer has the potential to improve RT target and organ segmentations, treatment planning, reduce RT-related complications, and ultimately enhance the quality of life for female patients undergoing pelvic RT cancer treatment.

\section{Acknowledgment}
This work is supported in part by the National Institutes of Health under award numbers DP5OD031876, R01CA272991, R01DE033512 and P30CA196521. 

\bibliographystyle{IEEEtran}  
\bibliography{main}  

\begin{thebibliography}{10}
\providecommand{\url}[1]{#1}
\csname url@samestyle\endcsname
\providecommand{\newblock}{\relax}
\providecommand{\bibinfo}[2]{#2}
\providecommand{\BIBentrySTDinterwordspacing}{\spaceskip=0pt\relax}
\providecommand{\BIBentryALTinterwordstretchfactor}{4}
\providecommand{\BIBentryALTinterwordspacing}{\spaceskip=\fontdimen2\font plus
\BIBentryALTinterwordstretchfactor\fontdimen3\font minus \fontdimen4\font\relax}
\providecommand{\BIBforeignlanguage}[2]{{%
\expandafter\ifx\csname l@#1\endcsname\relax
\typeout{** WARNING: IEEEtran.bst: No hyphenation pattern has been}%
\typeout{** loaded for the language `#1'. Using the pattern for}%
\typeout{** the default language instead.}%
\else
\language=\csname l@#1\endcsname
\fi
#2}}
\providecommand{\BIBdecl}{\relax}
\BIBdecl

\bibitem{bray2024global}
F.~Bray, M.~Laversanne, H.~Sung, J.~Ferlay, R.~L. Siegel, I.~Soerjomataram, and A.~Jemal, ``Global cancer statistics 2022: Globocan estimates of incidence and mortality worldwide for 36 cancers in 185 countries,'' \emph{CA: a cancer journal for clinicians}, vol.~74, no.~3, pp. 229--263, 2024.

\bibitem{marshall2022female}
D.~C. Marshall, E.~S. Tarras, A.~Ali, J.~Bloom, M.~A. Torres, and J.~M. Kahn, ``Female erectile tissues and sexual dysfunction after pelvic radiotherapy: A scoping review,'' \emph{CA: a cancer journal for clinicians}, vol.~72, no.~4, pp. 353--359, 2022.

\bibitem{basch2023patient}
E.~Basch, A.~C. Dueck, S.~A. Mitchell, H.~Mamon, M.~Weiser, L.~Saltz, M.~Gollub, L.~Rogak, B.~Ginos, G.~L. Mazza \emph{et~al.}, ``Patient-reported outcomes during and after treatment for locally advanced rectal cancer in the prospect trial (alliance n1048),'' \emph{Journal of Clinical Oncology}, vol.~41, no.~21, pp. 3724--3734, 2023.

\bibitem{lindau2015manifesto}
S.~T. Lindau, E.~M. Abramsohn, and A.~C. Matthews, ``A manifesto on the preservation of sexual function in women and girls with cancer,'' \emph{American Journal of Obstetrics and Gynecology}, vol. 213, no.~2, pp. 166--174, 2015.

\bibitem{unger2022sex}
J.~M. Unger, R.~Vaidya, K.~S. Albain, M.~LeBlanc, L.~M. Minasian, C.~C. Gotay, N.~L. Henry, M.~J. Fisch, S.~M. Lee, C.~D. Blanke \emph{et~al.}, ``Sex differences in risk of severe adverse events in patients receiving immunotherapy, targeted therapy, or chemotherapy in cancer clinical trials,'' \emph{Journal of Clinical Oncology}, vol.~40, no.~13, pp. 1474--1486, 2022.

\bibitem{peters2023clinical}
B.~Peters, A.~Ndumele, and M.~I. Uloko, ``Clinical implications of the historical, medical, and social neglect of the clitoris,'' \emph{The journal of sexual medicine}, vol.~20, no.~4, pp. 418--421, 2023.

\bibitem{jannini2014beyond}
E.~A. Jannini, O.~Buisson, and A.~Rubio-Casillas, ``Beyond the g-spot: clitourethrovaginal complex anatomy in female orgasm,'' \emph{Nature Reviews Urology}, vol.~11, no.~9, pp. 531--538, 2014.

\bibitem{di2014anatomic}
V.~Di~Marino, H.~Lepidi \emph{et~al.}, \emph{Anatomic study of the clitoris and the bulbo-clitoral organ}.\hskip 1em plus 0.5em minus 0.4em\relax Springer, 2014, vol.~91.

\bibitem{doo2025sex}
F.~Doo, W.~Naranjo, T.~Kapouranis, M.~Thor, M.~Chao, X.~Yang, and D.~Marshall, ``Sex-based bias in artificial intelligence-based segmentation models in clinical oncology,'' \emph{Clinical Oncology}, vol.~39, p. 103758, 2025.

\bibitem{ronneberger2015u}
O.~Ronneberger, P.~Fischer, and T.~Brox, ``U-net: Convolutional networks for biomedical image segmentation,'' in \emph{Medical image computing and computer-assisted intervention--MICCAI 2015: 18th international conference, Munich, Germany, October 5-9, 2015, proceedings, part III 18}.\hskip 1em plus 0.5em minus 0.4em\relax Springer, 2015, pp. 234--241.

\bibitem{isensee2021nnu}
F.~Isensee, P.~F. Jaeger, S.~A. Kohl, J.~Petersen, and K.~H. Maier-Hein, ``nnu-net: a self-configuring method for deep learning-based biomedical image segmentation,'' \emph{Nature methods}, vol.~18, no.~2, pp. 203--211, 2021.

\bibitem{hatamizadeh2021swin}
A.~Hatamizadeh, V.~Nath, Y.~Tang, D.~Yang, H.~R. Roth, and D.~Xu, ``Swin unetr: Swin transformers for semantic segmentation of brain tumors in mri images,'' in \emph{International MICCAI brainlesion workshop}.\hskip 1em plus 0.5em minus 0.4em\relax Springer, 2021, pp. 272--284.

\bibitem{chen2021transunet}
J.~Chen, Y.~Lu, Q.~Yu, X.~Luo, E.~Adeli, Y.~Wang, L.~Lu, A.~L. Yuille, and Y.~Zhou, ``Transunet: Transformers make strong encoders for medical image segmentation,'' \emph{arXiv preprint arXiv:2102.04306}, 2021.

\bibitem{wang2025triad}
S.~Wang, M.~Safari, Q.~Li, C.-W. Chang, R.~L. Qiu, J.~Roper, D.~S. Yu, and X.~Yang, ``Triad: Vision foundation model for 3d magnetic resonance imaging,'' \emph{arXiv preprint arXiv:2502.14064}, 2025.

\bibitem{ma2024segment}
J.~Ma, Y.~He, F.~Li, L.~Han, C.~You, and B.~Wang, ``Segment anything in medical images,'' \emph{Nature Communications}, vol.~15, no.~1, p. 654, 2024.

\bibitem{wang2022medclip}
Z.~Wang, Z.~Wu, D.~Agarwal, and J.~Sun, ``Medclip: Contrastive learning from unpaired medical images and text,'' in \emph{Proceedings of the Conference on Empirical Methods in Natural Language Processing. Conference on Empirical Methods in Natural Language Processing}, vol. 2022, 2022, p. 3876.

\bibitem{li2025automatic}
Y.~Li, J.~F. Wynne, Y.~Wu, R.~L. Qiu, S.~Tian, T.~Wang, P.~R. Patel, D.~S. Yu, and X.~Yang, ``Automatic medical imaging segmentation via self-supervising large-scale convolutional neural networks,'' \emph{Radiotherapy and Oncology}, vol. 204, p. 110711, 2025.

\bibitem{wasserthal2023totalsegmentator}
J.~Wasserthal, H.-C. Breit, M.~T. Meyer, M.~Pradella, D.~Hinck, A.~W. Sauter, T.~Heye, D.~T. Boll, J.~Cyriac, S.~Yang \emph{et~al.}, ``Totalsegmentator: robust segmentation of 104 anatomic structures in ct images,'' \emph{Radiology: Artificial Intelligence}, vol.~5, no.~5, p. e230024, 2023.

\bibitem{ji2022amos}
Y.~Ji, H.~Bai, C.~Ge, J.~Yang, Y.~Zhu, R.~Zhang, Z.~Li, L.~Zhanng, W.~Ma, X.~Wan \emph{et~al.}, ``Amos: A large-scale abdominal multi-organ benchmark for versatile medical image segmentation,'' \emph{Advances in neural information processing systems}, vol.~35, pp. 36\,722--36\,732, 2022.

\bibitem{heo2024rotary}
B.~Heo, S.~Park, D.~Han, and S.~Yun, ``Rotary position embedding for vision transformer,'' in \emph{European Conference on Computer Vision}.\hskip 1em plus 0.5em minus 0.4em\relax Springer, 2024, pp. 289--305.

\bibitem{grattafiori2024llama}
A.~Grattafiori, A.~Dubey, A.~Jauhri, A.~Pandey, A.~Kadian, A.~Al-Dahle, A.~Letman, A.~Mathur, A.~Schelten, A.~Vaughan \emph{et~al.}, ``The llama 3 herd of models,'' \emph{arXiv preprint arXiv:2407.21783}, 2024.

\bibitem{ma2022fast}
J.~Ma, Y.~Zhang, S.~Gu, X.~An, Z.~Wang, C.~Ge, C.~Wang, F.~Zhang, Y.~Wang, Y.~Xu \emph{et~al.}, ``Fast and low-gpu-memory abdomen ct organ segmentation: the flare challenge,'' \emph{Medical Image Analysis}, vol.~82, p. 102616, 2022.

\bibitem{grossberg2017data}
A.~Grossberg, A.~Mohamed, H.~Elhalawani, W.~Bennett, K.~Smith, T.~Nolan, S.~Chamchod, M.~Kantor, T.~Browne, K.~Hutcheson \emph{et~al.}, ``Data from head and neck cancer ct atlas,'' \emph{The Cancer Imaging Archive}, vol.~10, p.~K9, 2017.

\bibitem{jin2020deep}
L.~Jin, J.~Yang, K.~Kuang, B.~Ni, Y.~Gao, Y.~Sun, P.~Gao, W.~Ma, M.~Tan, H.~Kang \emph{et~al.}, ``Deep-learning-assisted detection and segmentation of rib fractures from ct scans: Development and validation of fracnet,'' \emph{EBioMedicine}, vol.~62, 2020.

\bibitem{smith2015data}
K.~Smith, K.~Clark, W.~Bennett, T.~Nolan, J.~Kirby, M.~Wolfsberger, J.~Moulton, B.~Vendt, and J.~Freymann, ``Data from ct\_colonography,'' \emph{Cancer Imaging Arch}, vol.~10, p.~K9, 2015.

\bibitem{an2020ct}
P.~An, S.~Xu, S.~Harmon, E.~Turkbey, T.~Sanford, A.~Amalou, M.~Kassin, N.~Varble, M.~Blain, V.~Anderson \emph{et~al.}, ``Ct images in covid-19 [data set], the cancer imaging archive,'' \emph{Mode of access: https://doi. org/10.7937/tcia}, 2020.

\bibitem{huang2023stu}
Z.~Huang, H.~Wang, Z.~Deng, J.~Ye, Y.~Su, H.~Sun, J.~He, Y.~Gu, L.~Gu, S.~Zhang \emph{et~al.}, ``Stu-net: Scalable and transferable medical image segmentation models empowered by large-scale supervised pre-training,'' \emph{arXiv preprint arXiv:2304.06716}, 2023.

\bibitem{hamamci2024developing}
I.~E. Hamamci, S.~Er, F.~Almas, A.~G. Simsek, S.~N. Esirgun, I.~Dogan, M.~F. Dasdelen, O.~F. Durugol, B.~Wittmann, T.~Amiranashvili \emph{et~al.}, ``Developing generalist foundation models from a multimodal dataset for 3d computed tomography,'' \emph{arXiv preprint arXiv:2403.17834}, 2024.

\bibitem{stock2024generalist}
R.~Stock, S.~Denner, Y.~Kirchhoff, C.~Ulrich, M.~R. Rokuss, S.~Roy, N.~Disch, and K.~Maier-Hein, ``From generalist to specialist: Incorporating domain-knowledge into flamingo for chest x-ray report generation,'' in \emph{Medical Imaging with Deep Learning}, 2024.

\bibitem{darcet2025cluster}
T.~Darcet, F.~Baldassarre, M.~Oquab, J.~Mairal, and P.~Bojanowski, ``Cluster and predict latents patches for improved masked image modeling,'' \emph{arXiv preprint arXiv:2502.08769}, 2025.

\bibitem{oquab2023dinov2}
M.~Oquab, T.~Darcet, T.~Moutakanni, H.~Vo, M.~Szafraniec, V.~Khalidov, P.~Fernandez, D.~Haziza, F.~Massa, A.~El-Nouby \emph{et~al.}, ``Dinov2: Learning robust visual features without supervision,'' \emph{arXiv preprint arXiv:2304.07193}, 2023.

\bibitem{peng2022beit}
Z.~Peng, L.~Dong, H.~Bao, Q.~Ye, and F.~Wei, ``Beit v2: Masked image modeling with vector-quantized visual tokenizers,'' \emph{arXiv preprint arXiv:2208.06366}, 2022.

\bibitem{hatamizadeh2022unetr}
A.~Hatamizadeh, Y.~Tang, V.~Nath, D.~Yang, A.~Myronenko, B.~Landman, H.~R. Roth, and D.~Xu, ``Unetr: Transformers for 3d medical image segmentation,'' in \emph{Proceedings of the IEEE/CVF winter conference on applications of computer vision}, 2022, pp. 574--584.

\bibitem{wald2025primus}
T.~Wald, S.~Roy, F.~Isensee, C.~Ulrich, S.~Ziegler, D.~Trofimova, R.~Stock, M.~Baumgartner, G.~K{\"o}hler, and K.~Maier-Hein, ``Primus: Enforcing attention usage for 3d medical image segmentation,'' \emph{arXiv preprint arXiv:2503.01835}, 2025.

\bibitem{li2025towards}
Y.~Li, Y.~Lai, M.~Thor, D.~Marshall, Z.~Buchwald, D.~S. Yu, and X.~Yang, ``Towards universal text-driven ct image segmentation,'' \emph{arXiv preprint arXiv:2503.06030}, 2025.

\bibitem{hu2024advancing}
M.~Hu, J.~Qian, S.~Pan, Y.~Li, R.~L. Qiu, and X.~Yang, ``Advancing medical imaging with language models: featuring a spotlight on chatgpt,'' \emph{Physics in Medicine \& Biology}, vol.~69, no.~10, p. 10TR01, 2024.

\end{thebibliography}
\end{document}